# High transport critical current density obtained for Powder-In-Tube-processed MgB$_2$ tapes and wires using stainless steel and Cu-Ni tubes

H. Kumakura, A. Matsumoto, H. Fujii and K. Togano
National Institute for Materials Science
1-2-1, Sengen, Tsukuba, Ibaraki 305-0047, Japan

MgB$_2$ tapes and wires were fabricated by the Powder-In-Tube method. Stainless steel and Cu-Ni tubes were used as sheath materials, and no heat treatment was applied. The tapes made of stainless steel showed transport critical current density Jc of about 10,000A/cm$^2$ at 4.2K and 5T. A high Jc of about 300,000A/cm$^2$ was obtained by extrapolating the Jc-B curves to zero field. Multifilamentary(7-core) MgB$_2$ wire was successfully fabricated using Cu-Ni tubes. For both tapes and wires the grain connectivity of MgB$_2$ was as good as a high-pressure sintered bulk sample. However, the Jc of the Cu-Ni sheathed wire was lower than the stainless steel sheathed tape due to the lower packing density of MgB$_2$.

The recent discovery of the 39K superconductivity in MgB$_2$[1] aroused much interest from researchers, not only in basic physics, but also in the field of applied superconductivity. The much higher Tc of MgB$_2$ than that of conventional metallic superconductors let researchers expect that MgB$_2$ can be used at elevated temperatures as a conductor of a cryogen-free magnet. Experiments on MgB$_2$ bulks and tapes indicate that the MgB$_2$ system shows no weak coupling of grains and that grain alignment is not required to obtain large current transfer across the grains[2-4]. This is very advantageous compared to high-Tc oxide superconductors where grain alignment is essential to improve grain coupling and to obtain high transport Jc values. For high-Tc oxide superconductors, researchers have developed special techniques in order to improve grain connectivity. In the case of MgB$_2$, however, we can expect high transport currents without any special technique.

For practical applications of superconductors, such as magnets and cables, we have to develop tapes or wires. Several research groups have already started the fabrication of tapes and wires using MgB$_2$. Canfield et al. first fabricated dense MgB$_2$ wires by exposing tungsten-core boron filaments to Mg vapor and obtained good superconducting properties[5]. However, most MgB$_2$ tapes and wires are now fabricated by so-called Powder-In-Tube method[6-9]. MgB$_2$-reacted powder or a mixture of Mg and B powder with stoichiometric composition is packed into various metal tubes. The powder/metal composite tubes are cold-worked into tapes and wires, and these tapes and wires are then heat treated at 900-1000°C for several hours. Thus, we have to use sheath materials that are not reactive with Mg and B at this temperature range. The metal tubes we can use as sheath materials are limited, as indicated by Jin et al[7]. Recently, Grasso et al. fabricated tapes using MgB$_2$ powder and a Ag, Cu or Ni tube and obtained high transport currents without any heat treatment[10]. This process is very attractive from the aspect of practical applications because a fabrication process including no heat treatment leads to much reduction of fabrication costs of wires and tapes. Furthermore, we can use various metals as sheath materials. Because higher Jc is expected for a harder sheath material, we selected stainless steel and Cu-Ni tubes as sheath materials and fabricated MgB$_2$ tapes and wires.

Commercially available MgB$_2$ powder(Alfa-Aesar) was tightly packed into stainless steel tubes and Cu-10wt%Ni tubes of 7-10cm in length. This packing process was carried out in air. The inner and outer diameters of the tubes were 4mm and 6mm for stainless steel tubes and 10mm and 14mm for Cu-Ni tubes, respectively. These tubes were cold rolled into rectangular rods of about 2mm in size using groove rolling and then cold rolled into tapes. The final size of the tapes was about 4mm in width and about 0.5mm in thickness. The typical thickness of the MgB$_2$ layer was 0.25mm. Multifilamentary wires were fabricated using Cu-Ni tubes. Groove rolled rods made of Cu-Ni tubes were cold-drawn to wires with a diameter of 2mm, and seven wires were bundled and inserted into a Cu-Ni tube and cold-worked into wire again. The final diameter of this 7-core wire was 2mm. All the tapes and wires were cold-worked into final size without any breakage. Figure 1 shows the cross sections of stainless steel sheathed tape and Cu-Ni sheathed 7-core wire. For the scanning electron microscopy(SEM) and X-ray diffraction analysis rectangular samples were cut from the tapes and sheath materials were removed. The MgB$_2$ layer was rigid and the fractured cross section was shiny. Figures 2(a) and (b) show the fractured and polished cross sections respectively of the $MgB_2$ layer in the stainless steel sheathed tape. Densely stacked $MgB_2$ was observed in the fractured cross section. The grain size of $MgB_2$ after the cold rolling was sub-micron, which was comparable to the particle size of the starting powder, indicating that no fracture of $MgB_2$ grains occurred during the cold working. X-ray diffraction analysis of the $MgB_2$ layer indicates that the main phase was $MgB_2$ with random grain orientation. However, small peaks corresponding to MgO were observed.

These tapes and wires were cut into short pieces of 3-4cm in length, and critical current Ic was measured. Ic was measured by a four-probe resistive method at 4.2K in magnetic fields. Current leads and voltage taps were directly soldered to the sheath materials of the tapes and wires. A magnetic field was applied parallel to the tape surface. The criterion of Ic definition was $1\mu V/cm$. For current I < Ic no voltage appeared. However, above Ic a rapid increase of voltage was observed, indicating that the superconducting-to-normal transition was fairly sharp. Critical current density Jc was obtained by dividing Ic by the cross sectional area of the $MgB_2$ core. Figure 3 shows Jc vs. magnetic field curves of the mono-core $MgB_2$/(stainless steel) tape, mono-core $MgB_2$/(Cu-Ni) tape and 7-core $MgB_2$/(Cu-Ni) wire at 4.2K. For comparison the Jc-B curve at 4.2K of the $MgB_2$ bulk prepared by the high pressure sintering and measured by the magnetization method is also shown in the figure[4]. All the tapes and wire show the same field dependence of Jc, and this field dependence of Jc is exactly the same as that of the high-pressure sintered bulk. This suggests that the connectivity of $MgB_2$ grains in all the tapes and wire is as good as the grain connectivity in the high-pressure sintered sample. The Jc of 7-core $MgB_2$/(Cu-Ni) wire was somewhat lower than that of the $MgB_2$/(Cu-Ni) tape. This should be attributed to the irregular cross section and sausaging of the $MgB_2$ filaments in the 7-core wire. High Jc values were obtained for $MgB_2$/(stainless steel) tape. Jc values of $MgB_2$/(stainless steel) were much higher than those of the $MgB_2$/(Cu-Ni) tape and wire although the cross sectional area reduction of $MgB_2$/(stainless steel) tape by the cold working is smaller than that of the $MgB_2$/(Cu-Ni) tape and wire. This can be explained by the difference of packing density of $MgB_2$ between the two samples. Because of the higher hardness of the stainless steel, stress applied to the $MgB_2$ in the stainless steel sheath during the cold rolling was higher than that in the Cu-Ni sheath. Thus, higher packing density of $MgB_2$ was obtained for $MgB_2$/(stainless steel) tape. Jc in 5T of $MgB_2$/(stainless steel) tape was about $10,000A/cm^2$ which was one of the highest transport Jc values ever reported for $MgB_2$ tapes and wires. This Jc is equal to or higher than that of the high-pressure sintered bulk. Below 5T precise Jc measurement was difficult for $MgB_2$/(stainless steel) tape because the heat generation at the current contacts increased the temperature of the sample and the tape was sometimes burned out immediately after the applied current exceeded Ic. Therefore we extrapolated the Jc-B curve of the $MgB_2$/(stainless steel) tape to a lower field, referring to the Jc-B curves of the high-pressure sintered bulk sample. The extrapolation of the Jc-B curve suggests that Jc at 0T was as high as $300,000A/cm^2$. However, due to the poor Jc tolerance against magnetic field, the Jc values in magnetic fields of our $MgB_2$/(stainless steel) tapes are still below the practical level. Because the packing density of $MgB_2$ layer is already high as shown in Fig. 1, Jc enhancement by increasing the packing density is limited for our sample. Thus, introduction of pinning centers is required to obtain substantial increase of Jc. The introduction of pinning centers is also effective in reducing the sensitivity of Jc to a magnetic field.

In summary, we fabricated $MgB_2$/(stainless steel) and $MgB_2$/(Cu-Ni) tapes by the PIT method without any heat treatment. $MgB_2$ grain connectivity for both tapes was as good as that of the high-pressure sintered bulk. Jc of $10,000A/cm^2$ at 4.2K and 5T was obtained for the $MgB_2$/(stainless steel) tape. Extrapolation of the Jc-B curve suggests the high Jc of $\sim 30,0000A/cm^2$ at 4.2K and zero field. Such high Jc values can be explained by the high packing density of $MgB_2$ associated with the hard sheath material. $MgB_2$/(Cu-Ni) multifilamentary wires were also successfully fabricated by the PIT method without any heat treatment.

References
1. J. Nagamatsu, N. Nakagawa, Y. Zenitani and J. Akimitsu, Nature 410(2001)63.
2. D.C. Larbalestier, M.O. Rikel, L.D. Cooley, A.A. Polyanskii, J.Y. Jiang, S. Patnaik, X.Y. Cai, D.M. Feldmann, A. Gurevichi, A.A. Squitieri, M.T. Naus, C.B. Ecom, E.E. Hellstrom, R.J. Cava, K.A. Regan, N. Rogado, M.A. Hayward, T.He, J.S. Slusky, P. Khalifah, K. Inumaru and M. Haas, Nature 410(2001)186.
3. M. Kambara, N. HariBabu, E.S. Sadki, J.R. Cooper, H. Minami, D.A. Cardwell, A.M. Campbell and I.H. Inoue, Supercond. Sci. Technol. 14(2001)L5.
4. H. Kumakura, Y. Takano, H. Fujii and K. Togano, submitted to Physica C.
5. P.C. Canfield, D.K. Finnemore, S.L. Bud'ko, J.E.

Ostenson, G. Lapertot, C.E. Cunningham and C. Petrovic, cond-mat/0102289
6. B.A. Glowacki, M. Majoros, M. Vickers, J.E. Evetts, Y. Shi and I Mcdougall, Supercond. Sci. Technol. 14(2001)193.
7. S. Jin, H. Mavoori and R.B. van Dover, submitted to Nature.
8. M.D. Sumpton, X. Peng, E. Lee, M. Tomsic and E.W. Collings, cond-mat/0102441.
9. S. Soltanian, X.L. Wang, I. Kusevic, E. Babic, A.H. Li, H.K. Liu, E.W. Collings and S.X. Dou, cond-mat/0105152
10. G. Grasso, A. Malagoli, C. Ferdeghini, S. Roncallo, V. Braccini and A. S. Siri, cond-mat/0103563

Figure captions

Fig. 1. Cross sectional view of mono-core $MgB_2$ tape(stainless steel sheath) and 7-core $MgB_2$ wire(Cu-Ni sheath).

Fig. 2. Scanning electron micrographs of (a) fractured and (b) polished $MgB_2$ layers in stainless-steel-sheathed $MgB_2$ tape.

Fig. 3. Transport Jc vs. field curves at 4.2K of Ni-Cu sheathed mono-core tape, Cu-Ni sheathed 7-core wire and stainless-steel-sheathed mono-core tape. Jc values of high-pressure sintered bulk sample estimated by the magnetization method are also shown in the figure. Extrapolated Jc for the stainless steel sheathed tape is indicated by the dotted line.

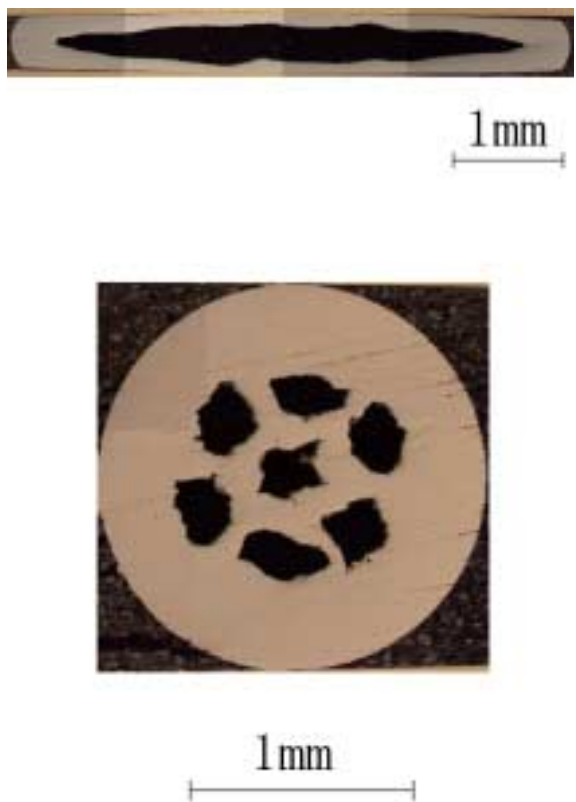

Fig. 1

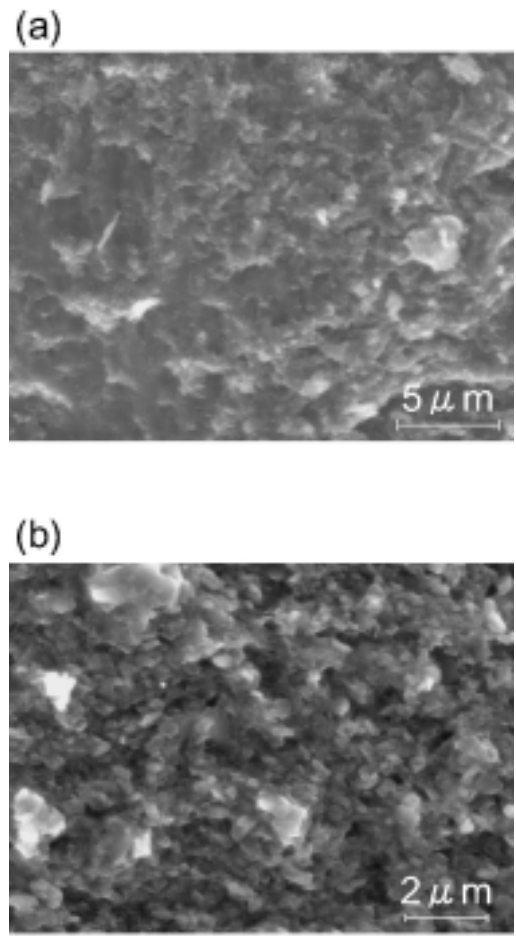

Fig. 2

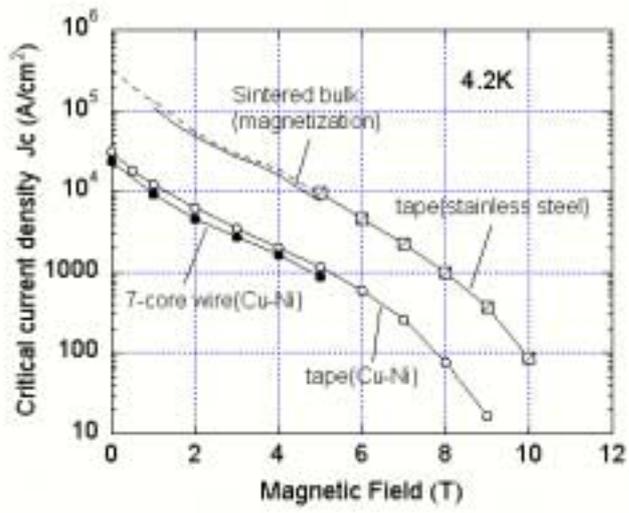

Fig. 3